\documentclass[a4paper,twoside,12pt]{article}
\input{obsjkt.sty}

\newcommand{\Msun}{\ensuremath{~{\rm M}_\odot}}                   
\newcommand{\Rsun}{\ensuremath{~{\rm R}_\odot}}                   
\newcommand{\Mjup}{\ensuremath{~{\rm M}_{\rm Jup}}}               
\newcommand{\Rjup}{\ensuremath{~{\rm R}_{\rm Jup}}}               
\newcommand{\rhosun}{\ensuremath{~\rho_\odot}}                    
\newcommand{\rhojup}{\ensuremath{~\rho_{\rm Jup}}}                
\newcommand{\Teff}{\ensuremath{T_{\rm eff}}}                      
\newcommand{\FeH}{\ensuremath{\rm [Fe/H]}}                        
\newcommand{\logg}{\ensuremath{\log g}}                           
\newcommand{\ms}{~m~s$^{-1}$}                                     
\newcommand{\mss}{~m~s$^{-2}$}                                    
\newcommand{\etal}{\textit{et al.}}                               

\newcommand{\tess}{\textit{TESS}}

\newcommand{\mcc}[1]{\multicolumn{3}{c}{#1}}

\newcommand{\ermcc}[5]{\mcc{\ensuremath{{#1\,^{+#2}_{-#3}}\,^{+#4}_{-#5}}}}

\begin{document} 

\OBSheader{The transiting planetary system WASP-86\,/\,KELT-12}{J.\ Southworth \& F.\ Faedi}{2022 Feb}

\OBStitle{The transiting planetary system WASP-86\,/\,KELT-12: \\ TESS provides the casting vote}

\OBSauth{John Southworth\,$^1$ and Francesca Faedi\,$^{2,3}$}

\OBSinst{Astrophysics Group, Keele University, Staffordshire, ST5 5BG, UK}
\OBSinst{Universit\`a degli Studi di Urbino ``Carlo Bo'', I-61029 Urbino, Italy}
\OBSinst{INFN, Sezione di Firenze, I-50019 Sesto Fiorentino, Firenze, Italy}

\OBSabstract{A transiting planetary system was discovered independently by two groups, under the names WASP-86 (Faedi \etal\ \cite{Faedi+16}) and KELT-12 (Stevens \etal\ \cite{Stevens+17aj}). The properties of the system determined in these works were very different, most tellingly a variation of a factor of three in the measured radius of the planet. We suggest that the system be named WASP-86\,/\,KELT-12 to better apportion the credit for discovery between the two groups. We analyse the light curve of this system from the Transiting Exoplanet Survey Satellite, which observed it in two sectors, following the \emph{Homogeneous Studies} approach. We find properties intermediate between the two previous studies: the star has a mass of $1.278 \pm 0.039$\Msun\ and a radius of $2.02 \pm 0.12$\Rsun, and the planet has a mass of $0.833 \pm 0.049$\Mjup\ and a radius of $1.382 \pm 0.089$\Rjup. The discrepancy in the two previous sets of measured properties of the system arises from a disagreement over the transit depth and duration, caused by the transit being long and shallow so not well suited to follow-up photometry from ground-based telescopes. We also update the orbital ephemeris to aid future work on this system, which is a good candidate for characterising the atmosphere of a planet through transmission spectroscopy.}


\section*{Introduction}

Transiting extrasolar planetary systems (TEPs) are the only systems beyond our own for which we can precisely measure the radii of planets, and thus their surface gravities and average densities. The first was discovered in late 1999 \cite{Henry+00apj,Charbonneau+00apj} and over 4000 are currently known. Apart from the scientific prospects, the study of TEPs has the procedural advantage that thousands of stars can be searched simultaneously to find transits, using telescopes with large fields of view. The downside to this approach is that only a small fraction of planets are transiting, and that many transits turn out to be due to false positives rather than real planets. This makes it necessary to survey a large number of stars in order to find a useful number of TEPs.

The first large source of TEP detections was small ground-based telescopes such as TrES \cite{Alonso+04apj}, HAT \cite{Bakos+02pasp}, SuperWASP \cite{Pollacco+06pasp}, KELT \cite{Pepper+07pasp} and HATSouth \cite{Bakos+13pasp}. Although there were a variety of approaches, all of these projects converged on the idea of using small telescopes (often commercially-available telephoto lenses) to survey large fractions of the available sky \cite{Pepper++03aca}. An inevitable consequence was that some TEPs were independently detected by multiple groups, for example HAT-P-10 \cite{Bakos+09apj} is identical to WASP-11 \cite{West+09aa}. Although this might seem inefficient due to the duplication of effort, it does have two advantages: it holds information on the completeness of the surveys and it allows a cross-check on the reliability of the properties measured for the TEPs discovered (e.g.\ Ref.\ \cite{Me10mn}).

In this work we revisit a planetary system that was announced almost simultaneously by two groups, under the names WASP-86 (Faedi \etal\ \cite{Faedi+16}) and KELT-12 (Stevens \etal\ \cite{Stevens+17aj}). The SuperWASP paper\cite{Faedi+16} included a small amount of follow-up light curves which did not cover the end of the transit well, plus a set of high-precision radial velocities (RVs) which yielded a clear detection of the planet and a tentative detection of an additional trend likely due to a third body on a wider orbit. The KELT paper \cite{Stevens+17aj} presented extensive photometry from small telescopes which, when combined, fully covered the transit. High-resolution speckle and adaptive-optics imaging was also obtained, as well as high-precision RVs that confirmed the additional trend at the 2.4$\sigma$ level.

Our interest in this object was recently revived in the course of a study of the suitability of the known TEPs for atmospheric characterisation. The TEPCat database\footnote{TEPCat is the Transiting Extrasolar Planet Catalogue (Southworth \cite{Me11mn}) at \texttt{https://www.astro.keele.ac.uk/jkt/tepcat/}.} listed the object as WASP-86, giving priority to the SuperWASP announcement as it predates the KELT announcement\footnote{The discovery paper of WASP-86 is datestamped 16 August 2016 on the \texttt{arXiv.org} preprint server, whereas the KELT paper is datestamped 2 September 2016.} whereas the NASA Exoplanet Archive\footnote{\texttt{https://exoplanetarchive.ipac.caltech.edu}} lists the properties from KELT and omits any mention of WASP-86. We found WASP-86 to be a poor target, but KELT-12 to be a very good target, for atmospheric characterisation.

Closer inspection revealed a huge difference in the properties of the TEP in the two discovery papers, which we illustrate in Table\,\ref{tab:comp}. Although many quantities vary significantly, some in particular stand out. The transit depth (which depends on the ratio of the radii) and duration (which depends on the sum of the radii divided by the semimajor axis of the relative orbit) are very discrepant. This is due to the shallow and long transit which makes it poorly suited to photometry from ground-based telescopes, and implies that high-precision photometry from a space-based observatory would be helpful. The masses and radii of the star are rather different, although the \Teff\ and \FeH\ measurements are consistent. Most strikingly, the planetary radius measurements differ by almost a factor of three.

\begin{table}[t]
\caption{\em Physical properties of WASP-86\,/\,KELT-12 from the two discovery papers. Some quantities were
not given directly in the papers so have been calculated from other quantities which were. \label{tab:comp}}
\centering
\begin{tabular}{lcc}
{\em Property}                      & {\em Faedi et al.\ \cite{Faedi+16}}     & {\em Stevens et al.\ \cite{Stevens+17aj}}     \\[3pt]
Stellar mass ($\!\!$\Msun)          & $1.239 \pm 0.018$                       & $1.591^{+0.070}_{-0.093}$                     \\
Stellar radius ($\!\!$\Rsun)        & $1.291^{+0.014}_{-0.013}$               & $2.37 \pm 0.17$                               \\
Stellar \logg\ (c.g.s.)             & $4.309 \pm 0.006$                       & $3.889^{+0.051}_{-0.050}$                     \\
Stellar density ($\!\!$\rhosun)     & $0.57 \pm 0.01$                         & $0.119^{+0.025}_{-0.020}$                     \\
Stellar \Teff\ (K)                  & $6330 \pm 110$                          & $6279 \pm 51$                                 \\
Stellar \FeH\ (dex)                 & $+0.23 \pm 0.14$                        & $+0.190^{+0.084}_{-0.085}$                    \\[3pt]
Orbital period (d)                  & $5.031555 \pm 0.000002$                 & $5.031623^{+0.000032}_{-0.000031}$            \\
Velocity amplitude ($\!\!$\ms)      & $84 \pm 5$                              & $82 \pm 12$                                   \\
Orbital eccentricity                & 0.0 fixed                               & 0.0 fixed                                     \\
Ratio of the radii                  & $0.0503 \pm 0.0008$                     & $0.0772^{+0.0019}_{-0.0018}$                  \\
Fractional stellar radius           & $0.0973$                                & $0.165 \pm 0.010$                             \\[3pt]
Planet mass ($\!\!$\Mjup)           & $0.821 \pm 0.056$                       & $0.95 \pm 0.14$                               \\
Planet radius ($\!\!$\Rjup)         & $0.632^{+0.014}_{-0.013}$               & $1.78^{+0.17}_{-0.16}$                        \\
Surface gravity ($\!\!$\mss)        & $46.8^{+3.3}_{-3.1}$                    & $7.45^{+1.8}_{-1.5}$                          \\
Planet density ($\!\!$\rhojup)      & $3.24^{+0.31}_{-0.26}$                  & $0.158^{+0.054}_{-0.040}$                     \\
Equilibrium temperature (K)         & $1415 \pm 22$                           & $1800 \pm 57$                                 \\
Semimajor axis (au)                 & $0.0617 \pm 0.0005$                     & $0.06708^{+0.00097}_{-0.0013}$                \\
\end{tabular}
\end{table}

We have therefore used data from the NASA Transiting Exoplanet Survey Satellite (\tess; Ref.\ \cite{Ricker+15jatis}) to establish robust properties for this system. For the rest of the current work we refer to the object under analysis as WASP-86\,/\,KELT-12 and suggest that this be adopted in the literature to ensure an equitable credit for the discovery of this planetary system.

Very little work has been published on WASP-86\,/\,KELT-12 since the discovery papers \cite{Faedi+16,Stevens+17aj}. Coker et al.\ \cite{Coker+18aj} presented high-resolution speckle imaging. They found no evidence for faint nearby companions, with limiting magnitude differences of 4.40 and 4.23 mag at 1.0$^{\prime\prime}$ separation, at wavelengths of 692 and 880 nm, respectively. The system is also listed under the name WASP-86 in the SWEET-Cat catalogue of stellar parameters\footnote{\texttt{http://sweetcat.iastro.pt}} (Santos et al.\ \cite{Santos+13aa}), where the host star is assigned the properties $\Teff = 6278 \pm 51$\,K, $\logg = 3.89 \pm 0.05$ and $\FeH = +0.19 \pm 0.08$.


\section*{Observational material}

\begin{figure}[t] \centering \includegraphics[width=\textwidth]{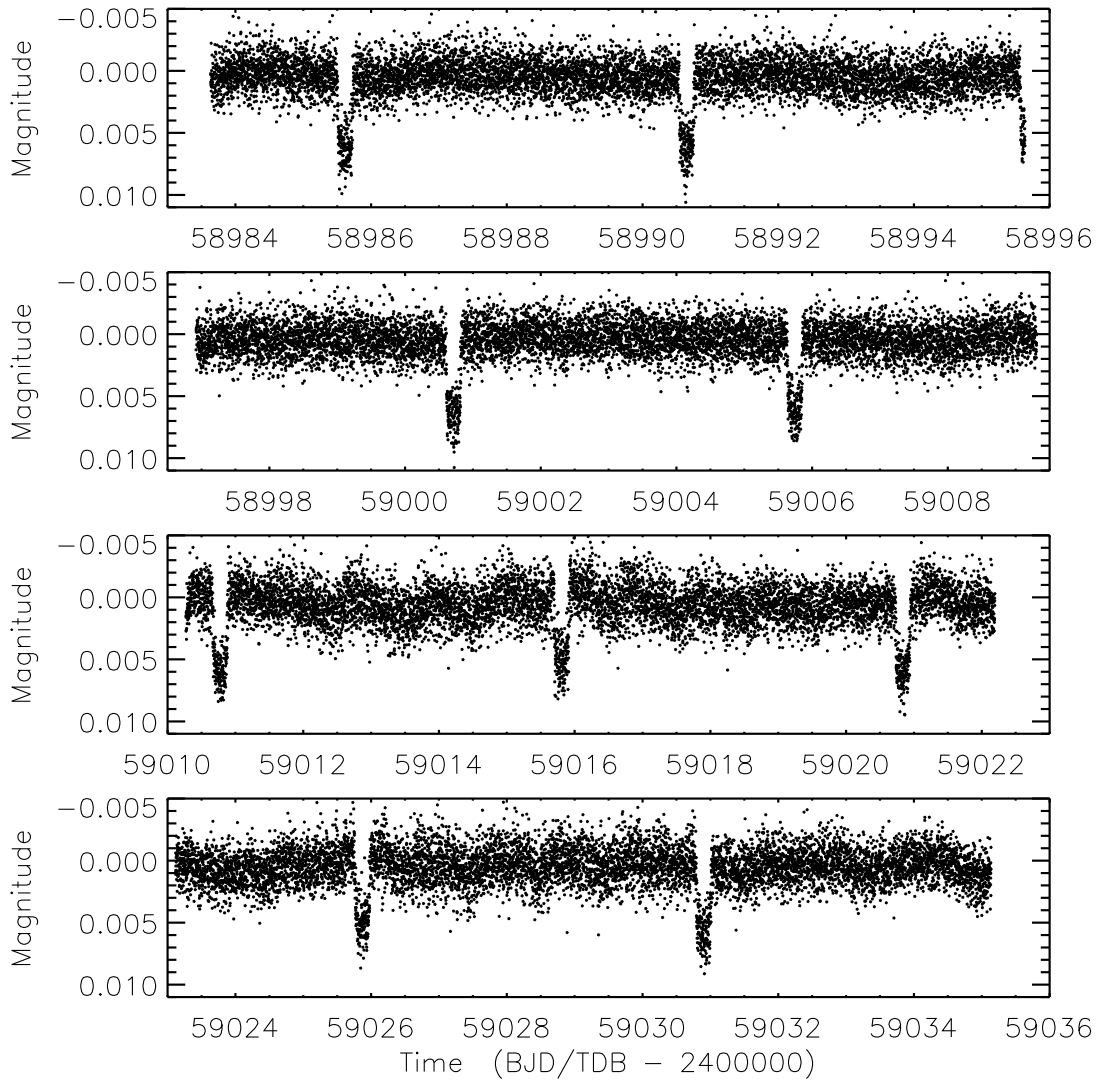} \\
\caption{\label{fig:time} \tess\ short-cadence PDCSAP photometry of WASP-86\,/\,KELT-12.
The top two panels show the data for sector 25 and the bottom two panels show the data for sector 26.}
\end{figure}


WASP-86\,/\,KELT-12 was observed using the NASA \tess\ satellite \cite{Ricker+15jatis} in sectors 25 (2020/05/13 to 2020/06/08) and 26 (2020/06/08 to 2020/07/04). The light curves comprise 18\,489 and 17\,909 datapoints obtained in short cadence mode \cite{Jenkins+16spie}, respectively, which were downloaded from the MAST archive\footnote{Mikulski Archive for Space Telescopes, \\ \texttt{https://mast.stsci.edu/portal/Mashup/Clients/Mast/Portal.html}} and converted to relative magnitude. All datapoints whose QUALITY flag was not zero were rejected, leaving a total of 34\,188 observations.

The \tess\ simple aperture photometry (SAP) had several instrumental trends which have been removed from the Pre-search Data Conditioning (PDCSAP) data, so we used the latter in our analysis (Fig.\,\ref{fig:time}). These data contain ten transits, of which one is only partially covered so was ignored. We selected the data during and close to each transit for further analysis, and fitted low-order polynomials to them to rectify them to zero differential magnitude. We rejected the data away from transit as they contain no useful information so merely slow down the computation of the best fit, leaving 4317 datapoints for our analysis.

Further data are planned\footnote{\texttt{https://heasarc.gsfc.nasa.gov/cgi-bin/tess/webtess/wtv.py?Entry=wasp-86}} to be obtained in \tess\ sectors 40 (2021/06/24 to 2021/07/23) and 52--53 (2022/05/18 to 2022/07/09). It will be worthwhile to revisit the system at a later data to include these data in an updated analysis.


\section*{Analysis of the \tess\ light curve}

\begin{figure}[t] \centering \includegraphics[width=\textwidth]{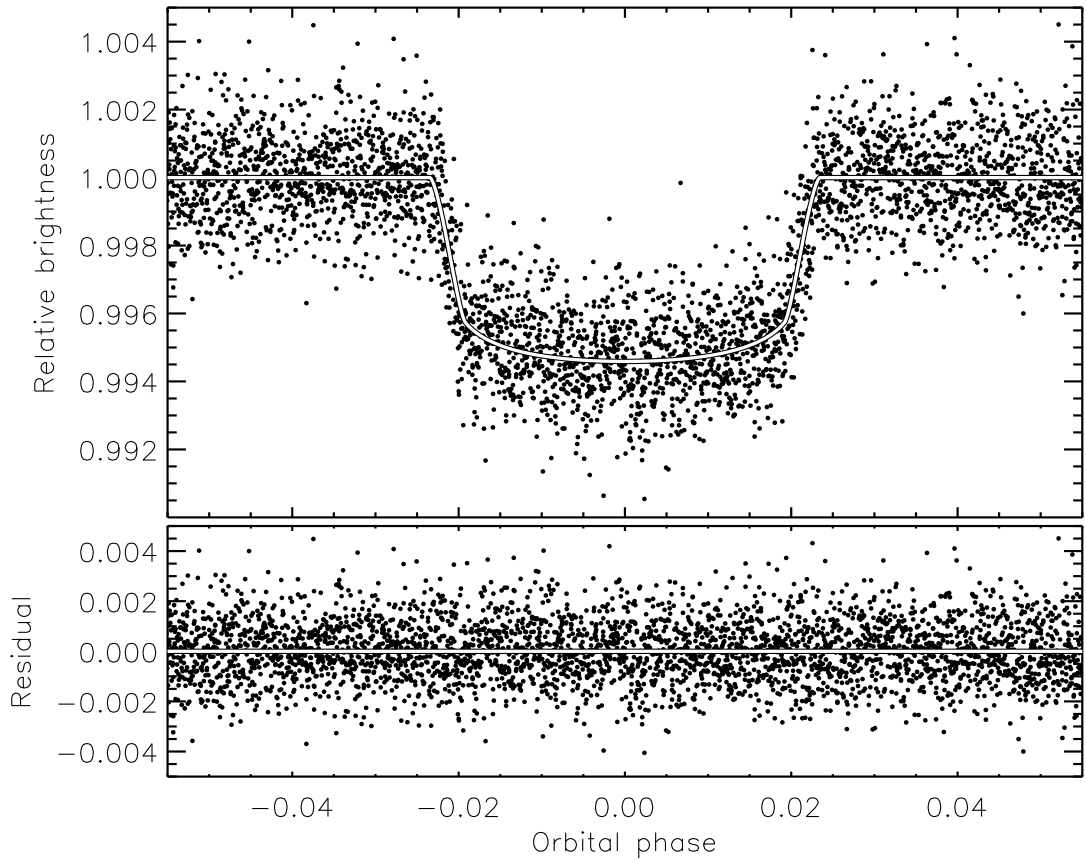} \\
\caption{\label{fig:tess} The \tess\ light curve of WASP-86\,/\,KELT-12
(filled circles) around the times of transit compared to the best fit (white
line). The residuals of the fit are shown in the lower panel.} \end{figure}

We analysed the \tess\ light curve of WASP-86\,/\,KELT-12 using the approach developed in the first author's \textit{Homogeneous Studies} papers (see Ref.\ \cite{Me08mn} and subsequent works). This is briefly described below. The errorbars were rescaled to force the reduced $\chi^2$ of the best fit to be unity.

The light curve was modelled using version 41 of the {\sc jktebop}\footnote{\texttt{http://www.astro.keele.ac.uk/jkt/codes/jktebop.html}} code \cite{Me++04mn2,Me13aa}. The fitted parameters were a reference time of mid-transit ($T_0$), the orbital period ($P$) and inclination ($i$), and the sum and ratio of the fractional radii ($r_{\rm A}+r_{\rm b}$ and $k = r_{\rm b}/r_{\rm A}$ where $r_{\rm A} = R_{\rm A}/a$ and $r_{\rm b} = R_{\rm b}/a$, $R_{\rm A}$ is the radius of the star, $R_{\rm b}$ is the radius of the planet and $a$ is the semi-major axis of the relative orbit). In the current work we subscript properties of the star with an `A' and of the planet with a `b'. We also fitted the coefficients of a straight line to normalise each transit to unit flux.

A circular orbit was assumed based on previous work \cite{Faedi+16,Stevens+17aj} and third light was assumed to be zero because the star appears to be isolated in the sky and no close companions have been found using high-resolution imaging \cite{Stevens+17aj,Coker+18aj}. The relatively modest upper limits from the speckle imaging are not problematic because Southworth \etal\ \cite{Me+20aa} found that any stars more than 3\,mag fainter than the planet host star have a negligible effect on the transit fit.

Limb darkening was included using four biparametric laws: quadratic, square-root, logarithmic and cubic \cite{Me08mn}. The data were fitted using two approaches for each law: both limb darkening coefficients fixed; and the linear coefficient fitted but the nonlinear coefficient fixed. There is no advantage in fitting for both limb darkening coefficients as they are strongly correlated \cite{Me++07aa} so primarily cause the minimisation process to be less stable. The values of the limb darkening coefficients were obtained from Claret \cite{Claret17aa} for solar metallicity. The best fit is shown in Fig.\,\ref{fig:tess}.

We specified as the reference time of minimum light one of the transit midpoints near the middle of the \tess\ dataset. The \tess\ data alone constrain the orbital period well, but we included the quoted time of inferior conjunction from Stevens \etal\ \cite{Stevens+17aj} to further improve the measurement. This time is 383 cycles earlier than our reference time and improves the precision of the period measurement by a factor of 55.

\begin{table}[t]
\caption{\em Results of the {\sc jktebop} analysis of the \tess\ light
curve of WASP-86\,/\,KELT-12. The errorbars are 1$\sigma$. \label{tab:lc}}
\centering
\begin{tabular}{llc}
{\em Quantity}                      & {\em Symbol}                & {\em Value}                 \\[3pt]
Orbital period (d)                  & $P$                         & $5.0316331 \pm 0.0000025$   \\
Time of minimum light (BJD/TDB)     & $T_0$                       & $2459010.77580 \pm 0.00035$ \\
Sum of the fractional radii         & $r_{\rm A}+r_{\rm b}$       & $0.1611 \pm 0.0086$         \\
Ratio of the radii                  & $k$                         & $0.07037 \pm 0.00081$       \\
Orbital inclination ($^\circ$)      & $i$                         & $85.9 \pm 1.2$              \\
Fractional radius of the star       & $r_{\rm A}$                 & $0.1505 \pm 0.0080$         \\
Fractional radius of the planet     & $r_{\rm b}$                 & $0.01059 \pm 0.00067$       \\
\end{tabular}
\end{table}

The uncertainties in the parameters of the fit were obtained in two ways: using Monte Carlo and residual-permutation algorithms \cite{Me08mn}. To this was added a contribution to the variation between fits with different treatment of limb darkening coefficients, which was small. The Monte Carlo errorbars are larger than the residual-permutation errorbars in this case, indicating that red noise in the \tess\ light curve is not significant. The fitted parameters and errorbars are given in Table\,\ref{tab:lc}.


\section*{Physical properties of WASP-86\,/\,KELT-12}

Although TEPs are a special case of eclipsing binary star system, they have the disadvantage that one piece of information is missing: the planet is not (normally) identifiable in spectra so its RVs are not measurable. An \emph{additional constraint} is needed, and is usually obtained by forcing the properties of the host star to match expectations for normal stars. This can be done using empirical calibrations of stellar properties \cite{Me10mn,Me09mn,Enoch+10aa} or by interpolating in the predictions of theoretical stellar models \cite{Me09mn,Sozzetti+07apj,Maxted++15aa}.

The theoretical-model approach is the more widely used because it yields high-precision results. The fractional radius of the star -- as measured from the transit light curve -- is very closely related to its density \cite{SeagerMallen03apj}. High-resolution spectroscopy can be used to determine the \Teff\ and \FeH\ of the star to high precision. Armed with these measurements, the mass, radius and age of the star follow from a comparison with theoretical predictions.

For the \Teff\ and \FeH\ of the host star in WASP-86\,/\,KELT-12 we used the values from Stevens et al.\ \cite{Stevens+17aj}. For the velocity amplitude of the star we took the weighted mean of the values from the two discovery papers (Table\,\ref{tab:comp}): $K_{\rm A} = 83.7 \pm 4.6$\ms.


\begin{table} \centering
\caption{\em Physical properties of WASP-86\,/\,KELT-12 obtained in this work. Where
one errorbar is given this is the random error. Where two sets of errorbars are given
the first is the random and the second is the systematic error. \label{tab:absdim}}
\begin{tabular}{lr@{\,$\pm$\,}c@{\,$\pm$\,}l}
{\em Parameter}                     & \multicolumn{3}{c}{\em Value}      \\[3pt]
Stellar mass ($\!\!$\Msun)          & 1.278    & 0.034    & 0.019        \\
Stellar radius ($\!\!$\Rsun)        & 2.02     & 0.12     & 0.01         \\
Stellar \logg\ (c.g.s.)             & 3.934    & 0.050    & 0.002        \\
Stellar density ($\!\!$\rhosun)     & \mcc{$0.155 \pm 0.027$}            \\[2pt]
Planet mass ($\!\!$\Mjup)           & 0.833    & 0.048    & 0.008        \\
Planet radius ($\!\!$\Rjup)         & 1.382    & 0.089    & 0.007        \\
Surface gravity ($\!\!$\mss)        & \mcc{$10.8 \pm  1.5$}              \\
Planet density ($\!\!$\rhojup)      & 0.295    & 0.059    & 0.001        \\[2pt]
Equilibrium temperature (K)         & \mcc{$1722 \pm   51$}              \\
Semimajor axis (au)                 & 0.06237  & 0.00054  & 0.00031      \\
Age of system (Gyr)                 & \ermcc{1.5}{0.4}{0.5}{0.4}{0.4}    \\[5pt]
\end{tabular}
\end{table}

We continued to follow the \textit{Homogeneous Studies} approach \cite{Me10mn} and use tabulated predictions from theoretical stellar evolutionary models. We first estimated an initial value of the velocity amplitude of the \emph{planet}, $K_{\rm b}$, and used that plus $K_{\rm A}$, $r_{\rm A}$, $r_{\rm b}$, $i$ and $P$ to determine the full properties of both components using standard formulae \cite{Hilditch01book}. We then iterated the value of $K_{\rm b}$ to find the best match between the measured $r_{\rm A}$ and \Teff\ and the values of $R_{\rm A}/a$ and \Teff\ obtained by interpolation in the theoretical models. This was done for a range of ages via a grid search to arrive at a single best set of physical properties for the system. Finally, this was performed for five different sets of theoretical models \cite{Me10mn} to obtain five different estimates of the system properties. The uncertainties in all input values were propagated by rerunning the analysis for every input parameter plus and minus its uncertainty. We also obtained a systematic error for each parameter, which we took to be the largest difference between the mean and individual values for each parameter across the results from the five different sets of theoretical models.

\begin{figure}[t] \centering \includegraphics[width=\textwidth]{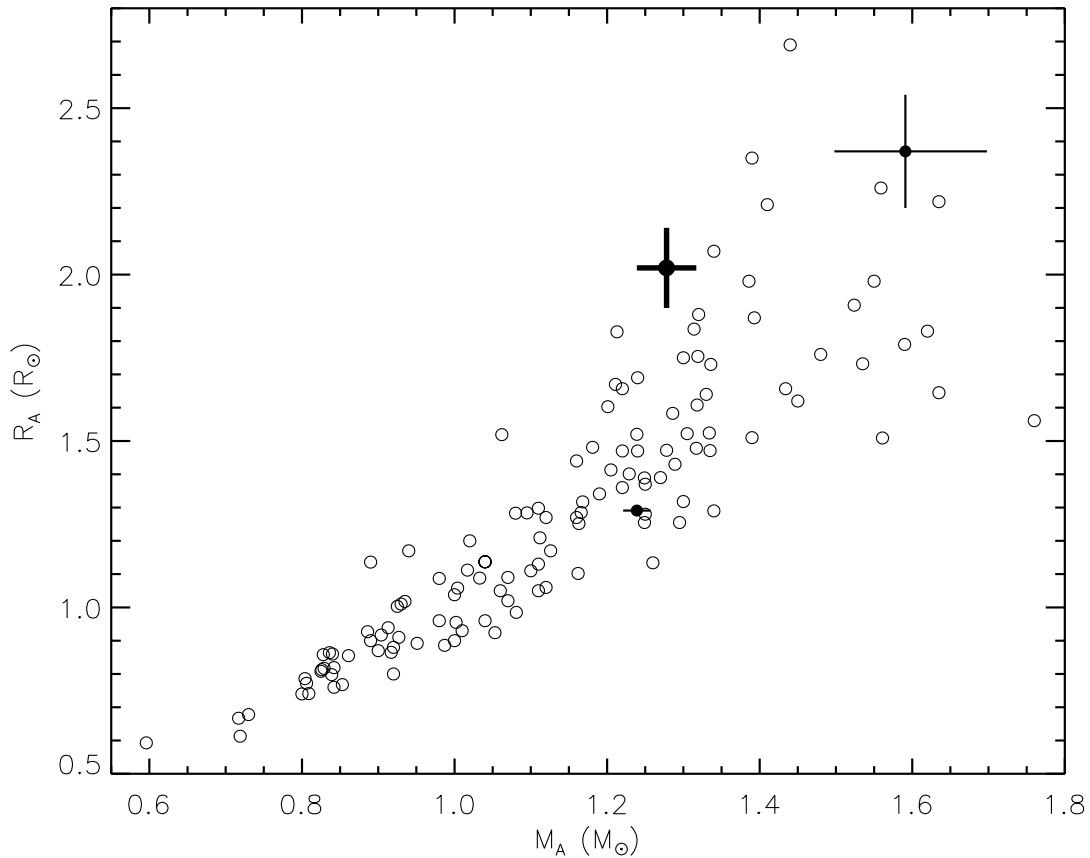} \\
\caption{\label{fig:mr1} Plot of the mass and radius measurements of the star in the
WASP-86\,/\,KELT-12 system. The results from Faedi \etal\ \cite{Faedi+16} and Stevens
et al.\ \cite{Stevens+17aj} are shown with filled points and errorbars. The measurements
from the current work are shown with a larger filled circle and thicker lines for the
errorbars. For context, the mass and radius measurements for the other planets discovered
by the SuperWASP and KELT consortia are shown with open circles (errorbars omitted for
clarity). Data taken from TEPCat \cite{Me15aspc} on 2021/08/24.}
\end{figure}

\begin{figure}[t] \centering \includegraphics[width=\textwidth]{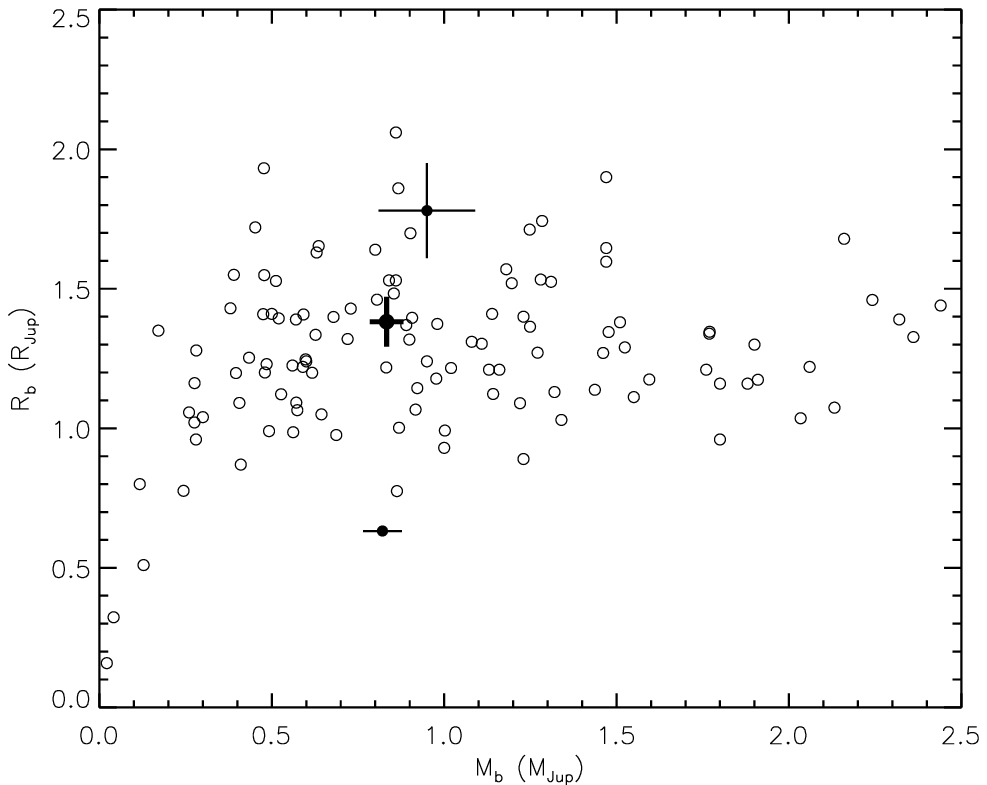} \\
\caption{\label{fig:mr2} Plot of the mass and radius measurements of the planet in
the WASP-86\,/\,KELT-12 system. Other comments are the same as for Fig.\,\ref{fig:mr1}.}
\end{figure}

The final parameters and uncertainties are given in Table\,\ref{tab:absdim}. Random and systematic errors are given for all measured quantities, with the exception of those which do not depend on stellar theory thus have no systematic error. A comparison between Tables \ref{tab:lc} and \ref{tab:absdim} and Table\,\ref{tab:comp} shows that our results are intermediate between those from the discovery papers, but are closer to those from Stevens \etal\ \cite{Stevens+17aj} than Faedi \etal\ \cite{Faedi+16}. This is further illustrated in Figs.\ \ref{fig:mr1} and \ref{fig:mr2}, which compare the sets of results in mass--radius diagrams for the two components of the system. We find the star to be moderately evolved and the planet to have an inflated radius as often seen in gas giants (e.g.\ \cite{Me+12mn3}).

It is clear that there are major differences in the three sets of physical properties measured for the WASP-86\,/\,KELT-12 system. We expect our new measurements to be the most reliable as they are based on much better light curves (courtesy of \tess) than previously available. Interestingly, the differences do not arise from spectroscopic measurements because the \Teff, \FeH\ and $K_{\rm A}$ values measured for the host star are very consistent across previous works \cite{Faedi+16,Stevens+17aj}. The discrepancy therefore must come from the density of the star, which is calculated almost directly from $r_{\rm A}$, which itself depends on the transit duration. Faedi \etal\ \cite{Faedi+16} significantly underestimated the transit duration, whereas Stevens \etal\ \cite{Stevens+17aj} slightly overestimated it. Both issues can be attributed to the difficulty of measuring the transit shape reliably using ground-based observations when the transit is this long (5.6 hours) and shallow (0.5\%). Stevens \etal\ did indeed note that their results ``can be heavily influenced by our choice of detrending parameters'' (see their section\,4.1).

In the discussion above we have assumed that the \tess\ light curve, and our model of it, are both reliable. Whilst this assumption appears safe, it is possible that a small amount of contaminating light exists which would have the effect of making the transit shallower and thus causing us to underestimate the radius of the planet. It is unfortunately not possible to determine the amount of contaminating light directly from the \tess\ light curve (see discussion in Southworth \cite{Me10mn}) so we are unable to remove this caveat from our analysis.


\section*{Summary}

The transiting planetary system WASP-86\,/\,KELT-12 was discovered independently by the SuperWASP \cite{Faedi+16} and KELT \cite{Stevens+17aj} groups, the two announcements occuring within a few weeks of each other. However, the properties of the system measured by the two groups were in poor agreement; the most obvious being a factor of three difference in the radius of the planetary component.

The main problem was the difficulty of obtaining suitable light curves of the transit, which is 5.6 hours long and only 0.5\% deep. We have therefore sought to establish reliable properties of the system using the light curve obtained by the \tess\ mission, which observed ten consecutive transits over the course of 51.5\,d. We used the \textit{Homogeneous Studies} approach to analyses these data: we first modelled the \tess\ light curve then deduced the properties of the system by requiring the host star to match tabulated predictions from several sets of theoretical stellar models.

We find system properties that are midway between those in the discovery papers \cite{Faedi+16,Stevens+17aj}. The SuperWASP analysis yielded a transit duration that was too small, leading to an overestimate of the stellar density and an underestimate of the radii of both components. The KELT analysis was closer to our results, but had a slightly over-long transit duration and thus an overestimate of the two radii. These issues underline the difficulty of obtaining good light curves of transits like these from ground-based telescopes. From our analysis of the \tess\ light curve we find that the host star is moderately evolved and that the planet is one of the class of inflated hot Jupiters \cite{Fortney++07apj}.

Other quasi-simultaneous independent discoveries have also occurred in the past, for example HAT-P-10 \cite{Bakos+09apj} and WASP-11 \cite{West+09aa}, HAT-P-27 \cite{Beky+12apj} and WASP-40 \cite{Anderson+11pasp}, MASCARA-2 \cite{Talens+18aa} and KELT-20 \cite{Lund+17aj}. Several independent announcements of new TEPs have also occurred based on separate groups following up the same \tess\ light curves. Whilst most sets of analyses are in mutual agreement, there are examples of strong disagreement on the physical properties (e.g.\ the subject of the current work) or even how many planets are needed to produce a set of observed transit events (e.g.\ TOI-561; Refs.\ \cite{Lacedelli+21mn,Weiss+21aj}). These instances can serve as informative cross-checks on the veracity of the properties measured for TEPs.


\section*{Acknowledgements}

We thank Jake Morgan for accidentally triggering this work, and Andrew Cameron and Don Pollacco for helpful comments.
This paper includes data collected by the \tess\ mission. Funding for the \tess\ mission is provided by the NASA's Science Mission Directorate.
The following resources were used in the course of this work: the NASA Astrophysics Data System; the SIMBAD database operated at CDS, Strasbourg, France; and the ar$\chi$iv scientific paper preprint service operated by Cornell University.



\end{document}